\documentclass[a4paper]{scrartcl}

\usepackage[affil-it]{authblk}

\usepackage{amsmath}
\usepackage{txfonts}
\usepackage{bm}
\usepackage{algorithm,algorithmicx,algpseudocode}

\usepackage{graphicx}

\newcommand{\functionname}[1]{\textsc{#1}}

\begin{document}

\title{Quicksort with median of medians is considered practical}
\author{Noriyuki \textsc{Kurosawa} \thanks{\texttt{kurosawa@vortex.c.u-tokyo.ac.jp}}}

\maketitle

\begin{abstract}
The linear pivot selection algorithm, known as median-of-medians, makes the worst case complexity of quicksort be \(\mathrm{O}(n\ln n)\). Nevertheless, it has often been said that this algorithm is too expensive to use in quicksort. In this article, we show that we can make the quicksort with this kind of pivot selection approach be efficient.
\end{abstract}


\section{Introduction}
Quicksort\cite{hoare_algorithm_1961-2,hoare_algorithm_1961-1,hoare_quicksort_1962} is one of the most efficient and widely used sorting algorithms.
The major drawback of quicksort is that worst case time complexity of naive implementations of quicksort is \(\mathrm{O}(n^2)\) with input size \(n\).
In order to avoid this worst case behavior, a variant of quicksort, so-called introsort\cite{musser_introspective_1997}
\footnote{
In many textbooks, it is written that this method is invented in ref.~\cite{musser_introspective_1997} in 1997.
We found, however, that the same technique has already appeared in ref.~\cite{ishihata_algorithm_1989}, which is published in 1989. In this book, the author calls this method \textit{fail-safe quicksort}.
}
, is often used.

There exists another solution to keep the worst case complexity of quicksort \(\mathrm{O}(n\ln n)\). This is the use of median-of-medians or Blum-Floyd-Pratt-Rivest-Tarjan (BFPRT) algorithm --- the pivot selection algorithm in the linear median finding algorithm\cite{blum_time_1973}. Despite the theoretical importance of this scheme, because of its somewhat large constant factor, this method have been considered impractical to use as a pivot selection algorithm in quicksort.

In this article, contrary to the above widespread view, we show that using BFPRT in quicksort is not inefficient, and moreover, considered practical.
Indeed, we find that quicksort with median-of-medians shows comparable performance to the well-known optimized quicksort implementations\cite{bentley_engineering_1993,durand_asymptotic_2003} for random sequence. We also show that the technique for improving performance is not only for the BFPRT but also available for other pivoting methods.

\section{Method}

\subsection{Idea}

The main idea of this article is thinning out the input elements of each pivot selection. The outline of this idea is given in Algorithm \ref{alg:select-pivot-bfprt}. Here we thin out the input array \(\bm{A}\) and apply the BFPRT on only \(1/s\) of \(\bm{A}\).
This BFPRT based pivot selection method guarantees the total time complexity of sorting to be \(\mathrm{O}(n \ln n)\), even in the case of \(s > 1\).
We can show this briefly as follows.
BFPRT guarantees that \(3/10\) of input elements are lesser than the output pivot, and other \(3/10\) are greater\cite{cormen_introduction_2009}. Thus, at least \(0.3n/s\) in \(\bm{A}\) with length \(n\) are lesser and other \(0.3n/s\) are greater. This yields that one partition operation reduces the length of \(\bm{A}\) from \(n\) to \([1-0.3/s]n\) even in the worst case. The total time complexity of sorting is then asymptotically \(\mathrm{O}(n\ln n)\) (see Ref.~\cite{cormen_introduction_2009}).

\begin{algorithm}
\caption{Pivot selection by BFPRT with thining out}
\label{alg:select-pivot-bfprt}
\begin{algorithmic}
\Function{select-pivot-with-thinning}{$\bm{A}$}
  \State Let $s$ be a parameter s.t. $s\ge 1$.
  \State $m \gets \functionname{length}(\bm{A})/s$.
  \State \Return (Apply BFPRT on $\left\{\bm{A}[0], \bm{A}[s], \bm{A}[2s], \dots, \bm{A}[(m-1)s]\right\}$).
\EndFunction
\end{algorithmic}
\end{algorithm}

The constant factor in the worst case of quicksort with the above method is estimated as follows. If the complexity follows the form \(a n\ln n+bn\) for sufficiently large \(n\), the recurrence relation becomes
\begin{align}
a n\ln n+ bn 
&= c_0 n + a (0.3\bar n)\ln(0.3\bar n) + b(0.3\bar n)
 +a (n-0.3\bar n)\ln\left(n-0.3\bar n) + b(n-0.3\bar n\right)
\nonumber\\
&=
a n \ln n + b n + \left[c_0 - a H\left(0.3/s\right)\right]n
,
\end{align}
here \(\bar n=n/s\) and \(H(p)=-p\ln p-(1-p)\ln(1-p)\) is the entropy function. The constant \(c_0\ge1\) represents the cost of the pivot selection and the partitioning.
If the bound of the complexity of BFPRT is \(c_1n\), \(c_0\) can be bound as \(c_0 \le 1.0+c_1/s\)
because the partitioning requires \(n-1\) comparisons\footnote{Because there are so many variants of BFPRT algorithms, we do not specify the value of \(c_1\) in this article.}.
The upper bound of \(a\) is then \((1.0+c_1/s)/H(0.3/s)\).
If \(s=40\) then \(a\lessapprox22.64+0.566c_1\), for example. The lower bound of \(a\) can be estimated in the same way: \(1.443(1.0+c_1/s)\lessapprox a\).

\subsection{Pseudo Median of \(3^L\)}

The idea of thinning out can be applicable for another pivot selection scheme, as long as it takes linear time.
For example, median-of-three\cite{singleton_algorithm_1969} method and median-of-three-medians-of-three (pseudo-median-of-nine or \textit{Tukey's ninther})\cite{bentley_engineering_1993,tukey_the_1978} are widely used pivot selection method. As the extreme case of these strategies, pseudo-median of $3^L$ ($L=\log_3 n$) can be considered.
Here we introduce another type of thinning method upon the pseudo-median of $3^L$, as described in Algorithm \ref{alg:select-pivot-pmed3L}.
\begin{algorithm}
\caption{Pivot selection by Pseudo median of $3^L$ with thining out}
\label{alg:select-pivot-pmed3L}
\begin{algorithmic}
\Function{select-pivot-by-pseudo-median}{$\bm{A}$}
  \State Let $s$ be a parameter s.t. $s\ge 1$.
  \State $n \gets \functionname{length}(\bm{A})$.
  \If{$n \le s$}
    \State \Return $\bm{A}[n/2]$ 
  \Else
    \State $a_0 \gets \functionname{select-pivot-by-pseudo-median}(\bm{A}[0..n/3-1])$
    \State $a_1 \gets \functionname{select-pivot-by-pseudo-median}(\bm{A}[n/3..2n/3-1])$
    \State $a_2 \gets \functionname{select-pivot-by-pseudo-median}(\bm{A}[2n/3..n-1])$
    \State \Return $\functionname{median-of-three}(a_0, a_1, a_2)$
  \EndIf
\EndFunction
\end{algorithmic}
\end{algorithm}

The worst case complexity of Alg.~\ref{alg:select-pivot-pmed3L} is estimated as the following.
If \(s=1\) and the length of the target array is \(3^L\), Alg.~\ref{alg:select-pivot-pmed3L} guarantees that at least \((2^L-1)/3^L\) of the elements are lesser than the selected pivot and other \((2^L-1)/3^L\) are greater. Thus, for a large array with length \(n\), Alg.~\ref{alg:select-pivot-pmed3L} guarantees that \(n^{\log_3 2-1}/s\) of the elements are greater than the pivot and \(n^{\log_3 2-1}/s\) are lesser. The depth of recursion \(d_{\text{r}}\) for the worst case is then estimated as
\begin{align}
d_{\text{r}} &\approx \log_{\left(1-n^{\log_3 2-1}/s\right)}(1/n)
\nonumber\\
&\simeq s n^{1-\log_3 2} \ln n \approx s n^{0.369}\ln n
.
\end{align}
Therefore, the worst case bound of quicksort with Alg.~\ref{alg:select-pivot-pmed3L} is bound by \(n d_{\text{r}} \approx sn^{1.369}\ln n \in \mathrm{O}(n^{1.369}\ln n)\).
Because the averaged required comparison in a pivot selection for an array with length \(n\) is \(1.333n\), the lower bound of the complexity of quicksort with this method in the best case is estimated as \(1.443(n+1.333n/s)\ln n = (1.443+1.924/s)n\ln n\).

\section{Experiment}

We carried out a numerical experiment to estimate the coefficients of the quicksort with Alg.~\ref{alg:select-pivot-bfprt} (t-BFPRT) and Alg.~\ref{alg:select-pivot-pmed3L} (t-PMed$3^L$). We used the standard single pivot partitioning\cite{hoare_algorithm_1961-2} and the original BFPRT algorithm\cite{blum_time_1973}. For simplicity, we did not change to another sorting algorithm such as heapsort or insertion sort to sort small subarrays. We also measured quicksort with random pivot picking (Hoare's original method) (Rand),  median-of-three (Med3), and pseudo-median-of-nine (PMed9) for benchmarking our strategy. We used random sequences of distinct integers as target arrays. We executed 100 times of sorting by each scheme, obtained the mean and unbiased variance of required count of comparisons, and calculated their coefficients via least squares fittings.

\begin{figure}
\centering
\includegraphics{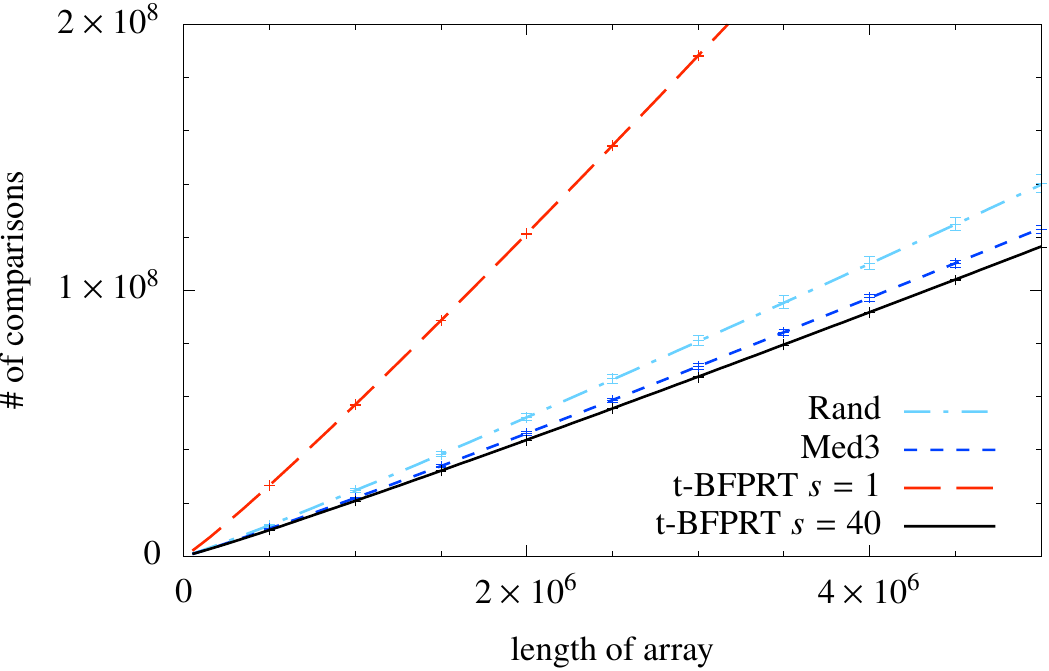}
\caption{Comparison number of t-BFPRT. We do not plot the line of PMed9 because it almost overlaps that of t-BFPRT(\(s=40\)).}
\label{fig:comparison-count-bfprt}
\end{figure}

\begin{figure}
\centering
\includegraphics{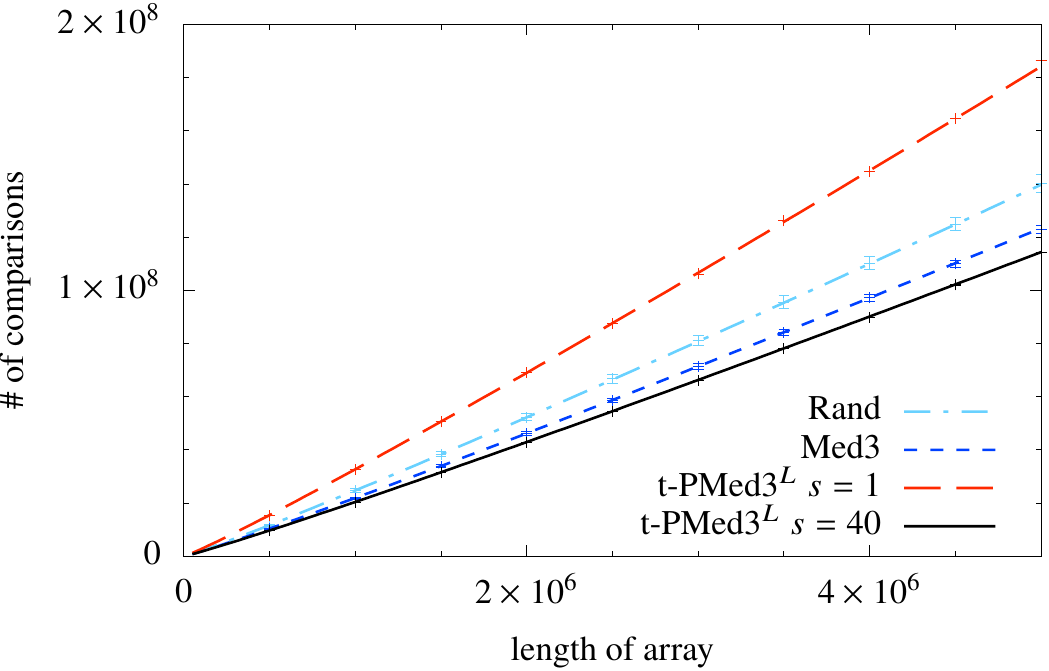}
\caption{Comparison number of t-PMed$3^L$. We do not plot the line of PMed9 because it almost overlaps that of t-PMed$3^L$(\(s=40\)).}
\label{fig:comparison-count-pmed3L}
\end{figure}

\begin{table}
\centering
\caption{Coefficients of each method for random sequences. (Fitting to \(an\ln n+bn\).)}
\label{tbl:coefficient-by-fitting}
\begin{tabular}{cl@{ $\pm$}ll@{ $\pm$}l}
\hline
method & \multicolumn{2}{c}{$a$} & \multicolumn{2}{c}{$b$}
\\
\hline
Rand & $2.004$ & $0.004$ & \phantom{1}$-2.89$ & $0.04$ 
\\
Med3 & $1.710$ & $0.002$ & \phantom{1}$-1.74$ & $0.02$
\\
PMed9 & $1.568$ & $0.001$ & \phantom{1}$-1.02$ & $0.01$
\\
t-BFPRT ($s=1$) & $5.224$ & $0.006$ & $-15.16$ & $0.09$
\\
t-BFPRT ($s=40$) & $1.5371$ & $0.0004$ & \phantom{1}$-0.455$ & $0.006$
\\
t-PMed$3^L$ ($s=3$) & $2.57$ & $0.03$ & \phantom{1}$-2.8$ & $0.4$
\\
t-PMed$3^L$ ($s=40$) & $1.528$ & $0.003$ & \phantom{1}$-0.72$ & $0.04$
\\
\hline
\end{tabular}
\end{table}

\begin{figure}
\centering
\includegraphics[width=0.3\textwidth]{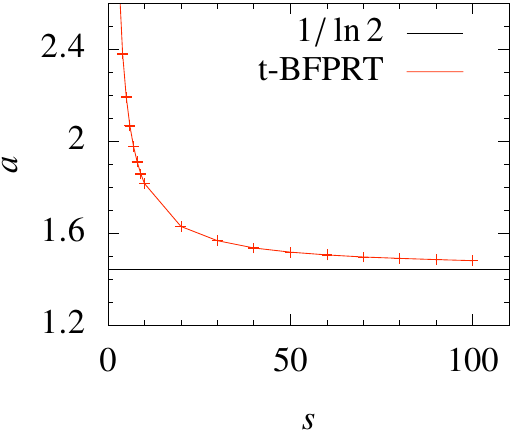}
\hspace{1em}
\includegraphics[width=0.3\textwidth]{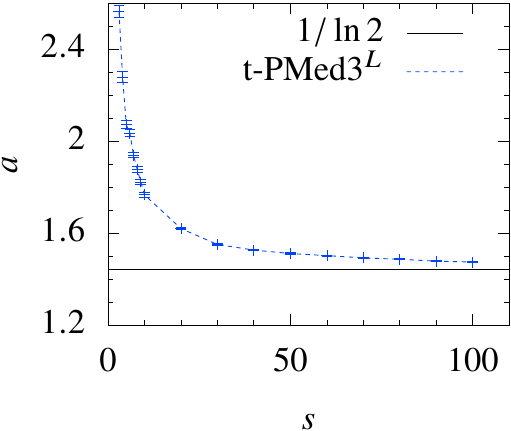}
\hspace{1em}
\includegraphics[width=0.3\textwidth]{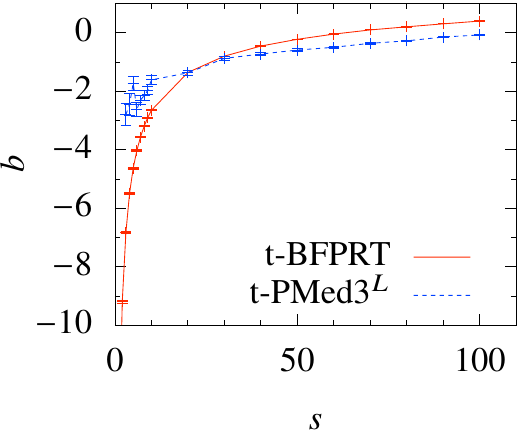}
\caption{Dependence of the coefficients on \(s\). We plot the line \(1/\ln 2\) as the theoretical limit.}
\label{fig:s-dependence}
\end{figure}

Figure \ref{fig:comparison-count-bfprt} and \ref{fig:comparison-count-pmed3L} show the number of comparison of each pivoting method.
We can see that the performance of t-BFPRT with \(s=1\) or t-PMed$3^L$ with \(s=3\) is not good because of its expensive pivot selection procedure. On the other hand, t-BFPRT or t-PMed$3^L$ with \(s=40\) exhibits excellent result; both method beat that of Med3.
Table \ref{tbl:coefficient-by-fitting} shows the fitting result to the function \(an\log n+bn\). We can see that t-BFPRT and t-PMed$3^L$ with large \(s\) clearly outperform Med3 and exhibit comparable result to PMed9.

Figure \ref{fig:s-dependence} shows the dependence of the coefficients on \(s\). In both methods, the larger \(s\) is, the more improved the asymptotic behavior is. Because the larger \(s\) makes the upper bound of the comparison count worse, \(s\) within the range \(20\lessapprox s \lessapprox 50\) is probably the best choice in this configuration. 

Note that the above discussion only considers the number of comparisons and the execution time is another question. As for t-PMed$3^L$, there is no swapping in pivot selection phase. Therefore, unless the effect of the memory cache and cost of function calls are not negligible, the execution time shows the same tendency of the comparison count. On the other hand, t-BFPRT involves data modifications in the pivot selection. In our environment, it showed the similar tendency of the comparison count. However, further accurate investigations should be required anyway.


\section{Conclusion}

In this article, we show that the use of median-of-medians as a pivot selection algorithm in quicksort is not always inefficient; it works very well on random sequences, but still requires only \(\mathrm{O}(n\ln n)\) comparisons in the worst case. We also show that the technique of thinning out can also be usable for other pivoting method, and can improve the worst case behavior with keeping good performance for random input. 

There can be many other methods to thinning out the input of pivoting algorithm.
 It is easy to make a bad input for Alg.~\ref{alg:select-pivot-bfprt} (setting larger elements at the multiple of \(5s\) and last \(n/s\) elements), but for another thinning method making a bad input may not be as easy as this. 
Also, the methods in this article perform well on neither increasing nor decreasing sequence because of its naive behavior for small subarrays. To fix this, equipping adaptive features (e.g., median-of-three for small input (typically \(n<5s\) for t-BFPRT and \(n<s\) for t-PMed$3^L$) and the main method for large input) is effective.

Because of its simple concept and efficiency, we believe that the presented approach can be useful in real world applications.


\end{document}